\documentclass[reqno]{amsart}
\usepackage{graphics,amsmath,xcolor}

\newcommand{\vf}{\mathfrak v} 
\newcommand{\ehc}{\Lambda} 
\newcommand{\cc}{{c}} 

\newcount\n  \newdimen\w

\def\Repeat#1#2{\n=#1\relax\loop\ifnum       
  \n>0\relax #2\advance\n by-1\repeat}

\long\def\OMIT#1{\relax }  

\def\re#1{(\ref{#1})}   
  
\def\eqn#1#2{ \begin{align} \label{#1}         #2 \end{align}}

\def\nl#1{          \\ \label{#1}        }  
\def\nnl#1{ \tag*{} \\ \label{#1}        }  

\def\delim#1#2#3{\csname\ifcase#1 relax\or   
   big\or Big\or bigg\or Bigg\fi\endcsname   
  {\ifcase#2\or\Delim#3\or\deliM#3\fi}}      
\def\Delim#1{\ifcase#1\relax\or(\or[\or\{\or<\or\langle\or|\or\|\or---{ }\fi}
\def\deliM#1{\ifcase#1\relax\or)\or]\or\}\or>\or\rangle\or|\or\|\or{ }---\fi}

\def\largerfrac#1#2#3{      
  \whichtypesize\n=\currenttypesize\advance\n by #1 \mathchoice
  {\setbox0\hbox{$\displaystyle-$} \w=.5\ht0\advance\w by-.5\dp0\setbox0
    \hbox{\typesize\n $\displaystyle-$} \advance\w by -.5\ht0\advance\w
    by .5\dp0\raise\w \hbox{\typesize\n$\displaystyle{\frac{#2}{#3}}$}}
  {\setbox0\hbox{$-$} \w=.5\ht0 \advance\w by -.5\dp0 \setbox0\hbox
    {\typesize\n $-$} \advance\w by-.5\ht0\advance\w by
    .5\dp0\raise\w\hbox{\typesize\n$\frac{#2}{#3}$}}
  {\setbox0\hbox{$\scriptstyle-$} \w=.5\ht0 \advance\w by-.5\dp0\setbox0
    \hbox{\typesize\n $\scriptstyle-$} \advance\w by -.5\ht0 \advance\w
    by .5\dp0 \raise\w\hbox{\typesize\n$\scriptstyle{\frac{#2}{#3}}$}}
  {\setbox0\hbox{$\scriptscriptstyle-$} \w=.5\ht0
    \advance\w by -.5\dp0 \setbox0\hbox{\typesize\n
    $\scriptscriptstyle-$} \advance\w by -.5\ht0 \advance\w by .5\dp0
    \raise\w\hbox{\typesize\n$\scriptscriptstyle{\frac{#2}{#3}}$}}  }

\begin{document}

\title{Spectral properties of dissipation}
\author{P. V\'an$^{1,2,3}$, R. Kov\'acs$^{1,2,3}$, and F. V\'azquez$^{4}$ }
\address{$^1$Department of Theoretical Physics, Wigner Research Centre for Physics, H-1525 Budapest, Konkoly-Thege Miklós u. 29-33., Hungary; //
$^2$Department of Energy Engineering, Faculty of Mechanical Engineering,  Budapest University of Technology and Economics, 1111 Budapest, Műegyetem rkp. 3., Hungary;//
$^{3}$ Montavid Thermodynamic Research Group, 1112 Budapest, Hungary;//
$^{4}$ Department of Physics, UAEM, Science Research Center, 62209 Cuernavaca, Av. Universidad 1001}
 
\date{\today}

\begin{abstract}
	The novel concept of {\em spectral diffusivity} is introduced to analyse {the} dissipative properties of continua. The dissipative components of a linear system of evolution equations are separated into noninteracting parts.  This separation is similar to mode analysis in wave propagation. The new modal quantities characterise dissipation and best interpreted as effective diffusivities, or, in case of heat conduction, as effective heat conductivities of the material. 
\end{abstract}

\maketitle

\section{Introduction}

Dispersion relations are useful  {since t}hey provide a particular insight of linear partial differential equations, identifying characteristic velocities, damping effects, dispersion properties and also linear stability conditions can be revealed with them \cite{Fut62a,Str05b,BerVan17b}. They are constructed when we are looking for a plane wave solution of the evolution of a field $\Phi$, in the following form 
\eqn{wavesol}{
	\Phi(x,t) = \Phi_0 e^{i(\omega t - k x)},
} 
where $\omega$ is the angular frequency, and $k$ is the wavenumber. Then a linear answer of a medium is characterised from the point of view of wave propagation. Then two different points of views are applied (see \cite{Str05b}). Either the wave number is considered real, and the angular frequency is a complex function,  {or} the angular frequency is considered real, and the wave number is a complex function. In the first case $\omega(k) = k v_p(k) + i \Omega(k)$, where the {\em phase velocity}, $v_p$, and the {\em damping factor}, $\Omega$, are real functions. In the following, it is called \emph{temporal representation}. Then one obtains
\eqn{wavesol1}{
	\Phi(x,t) = \Phi_0  e^{-\Omega t} e^{i k(v_p t - x)},
} 
where  $\Omega$ is expected to be nonnegative, otherwise the amplitude of plane wave solutions is increasing in time indicating a temporal instability of plane wave solutions. 

In the second case, when the angular frequency is considered real, a convenient  representation is $k(\omega) = \frac{\omega}{v_p(\omega)} - i \alpha(\omega)$, where  the phase velocity, $v_p(\omega)$, is a function of angular frequency and $\alpha$ is the {\em attenuation factor}. In the following, it is called \emph{spatial representation}. Then one obtains
\eqn{wavesol2}{
	\Phi(x,t) = \Phi_0  e^{-\alpha x} e^{i \omega(t - x/v_p)},
}
where $\alpha$ is expected to be nonnegative, otherwise the amplitude of plane wave solutions is increasing with  {distance,} and therefore spatial instability of plane wave solutions appear. This second representation is characteristic in acoustics and in seismology, where also the so-called {\em quality factor}, $Q = \frac{\omega}{2\alpha v_p}$ plays an important role as the parameter of dissipation. $Q$ is the relative energy loss in one period of the plane wave when compared to the maximum stored energy in the specimen. It is convenient in earth sciences because it is practically constant for earth materials in the typical seismological frequency range. 

All these concepts are based on aspects of wave propagation. The dissipation characteristics, the damping coefficient $\Omega$, the attenuation factor $\alpha$ {and the} quality factor $Q$ are related and defined with a plane wave solution in mind. In pure dissipative, diffusive systems, these concepts are somehow clumsy. Here we propose a new quantity, the {\em spectral diffusivity}, defined with the help of the damping and the attenuation factors respectively
\eqn{effhc}{
	\ehc(\omega) := Im \left(\frac{\omega}{k(\omega)^2}\right), \qquad \ehc(k) := Im\left(\frac{\omega(k)}{k^2}\right) = -\frac{\Omega}{k^2},
}
The two, wave number and angular frequency-dependent, spectral diffusivities are not the same, but similar, like the damping and attenuation. They represent temporal and spatial diffusive behaviour. 
 In the following, we will {argue} that this concept enables a more in-depth analysis of pure dissipative {systems} without wave-like solutions and also seems to be useful for analysing mixed {systems} with damped and attenuated waves. The apparent characteristic values are interpreted as channels of diffusive propagation like the various phase velocities {that} characterise waves. For complex systems, spectral diffusivities express effective properties, e.g. for heat conduction, these can be interpreted as effective thermal diffusivities.

In the following the equations are written in one spatial dimension. The wave equation for a scalar field, $\phi(t,x)$, reads as
\eqn{wave_eq}{
	\partial_{tt} \phi - \vf^2 \partial_{xx} \phi = 0,
}
where $\partial$ denotes the partial derivatives by time or position, according to the indices. $\vf$ is the wave velocity, characteristic to the medium.
The Fourier equation in one spatial dimension for the temperature field, $T(t,x)$, is given as 
\eqn{Fourier_eq}{
	\partial_{t} T - \lambda \partial_{xx} T = 0,
}
where $\lambda = \frac{\lambda_F}{\rho c}$ is the thermal diffusivity, that is the Fourier heat conduction coefficient, $\lambda_F$, divided by the density, $\rho$, and the specific heat, $c$. The wave solutions with the form \re{wavesol} are  obtained if the frequency $\omega$ and the wave number $k$ are not independent. The {\em dispersion relation} of the evolution equations, \re{wave_eq} and \re{Fourier_eq} are 
\eqn{w_rel}{
	\omega^2 -\vf^2 k^2 = 0 \qquad \text{and} \qquad
	i \omega + \lambda k^2 = 0,
}
respectively. For the wave equation the phase velocity is constant, $v_p = \vf$ and the dissipation parameters are zero, wave equation is related to an ideal continuum without dissipation from a thermodynamic point of view. For the Fourier equation the phase velocity is zero, the damping coefficient, $\Omega = \lambda k^2$. The other representation with real angular frequency, including the attenuation factor, does not seem to be really informative. However, the spectral diffusivity $\ehc(k)=\ehc(\omega) = \lambda$. 


In the following, we treat three simple examples of heat conduction. First, the Maxwell-Cattaneo-Vernotte equation is analysed. This is a telegraph type equation, where both diffusive and wave-like propagation is present. In the limit of zero dissipation appears pure wave propagation, and in the limit of zero inertia, one gets pure diffusive behaviour. Our second example is composed of two heat-conducting Fourier channels with heat exchange. It is a pure dissipative system, where the phase velocities are zero. It is shown, that here the spectral thermal diffusivity has two values, related to the particular channels of diffusive propagation. We will see that the heat exchange influences effective properties. Finally, a more complicated example is considered, where the analysis of the dissipation was crucial from an experimental point of view. We calculate the "window condition" in the 9-field theory of Extended Thermodynamics, where both second sound and ballistic propagation is present. {The} maximum of the spectral diffusivity determines the optimal frequency for second-sound observations, where dispersion and damping {are} minimal.

\section{Diffusion and wave propagation: Maxwell-Cattaneo-Vernotte (MCV) type heat conductors}

The dispersion relation is a solution of the linear or linearised partial differential equation of a continuum to particular initial conditions. For the wave equation in one spatial dimension, this is the d'Alembert solution, with two initial conditions, without boundary requirements. In general, wave propagation without dissipation may be rather complicated \cite{BerVan17b}. The simplest combination of ideal and dissipative phenomena, wave-like and diffusive propagation, is best analysed with the help of the MCV equation. It is a hyperbolic model of wave-like propagation of internal energy due to inertial effects in heat propagation. For the temperature, we obtain a second-order partial differential equation for the temperature, which is equivalent to the telegraph equation for damped electromagnetic waves \cite{MorFes53b}. In case of heat conduction, the Fourier like diffusive propagation is the primary process, wave-like propagation never appears alone. The observation of heat waves requires special conditions and specific materials \cite{Gya77a,JouAta92b,MulRug98b,JacWal71a}. The related equations are the balance of internal energy and the MCV evolution equation for the heat flux:
\eqn{iebal_eq}{
	\rho c \partial_t T + \partial_x q &= 0, \nl{MCV_eq}
	\tau \partial_t q + q + \lambda_F \partial_x T &= 0. 
} 

Here $\tau$ is the relaxation time, representing the inertial, memory effects in heat propagation. If $\tau$ is zero, then the system can be reduced to the Fourier equation \re{Fourier_eq}. The substitution of the wave solution for both variables, $(T,q)(x,t) = (T_0,q_0)e^{i (\omega t - k x)}$ results in the following matrix equation
\eqn{MCV_leq}{
	\begin{pmatrix}
		i \omega \rho c & - i k \\- i k \lambda_F & 1 +  i \omega \tau
	\end{pmatrix} 
	\begin{pmatrix}T_0 \\ q_0\end{pmatrix}
	= \begin{pmatrix}0 \\ 0\end{pmatrix}.
}
If the determinant of the matrix is zero, the nontrivial solutions of the inequality are determined by the dispersion relation:
\eqn{MCV_drel}{
	\tau \omega^2 - i \omega - \lambda k^2 = 0, \qquad \left(\lambda = \frac{\lambda_F}{\rho c}\right).
}
Then it is easy to calculate the spectral diffusivities in both the temporal, real $k$, and in the spatial, real $\omega$.  cases. For the temporal case the angular frequency is the following:
\eqn{MCV_drelo}{
	\omega_{MCV}(k) = \frac{i \pm \sqrt{4\tau\lambda k^2-1}}{2\tau}.
}
Therefore the spectral diffusivity becomes
\eqn{MCV_ehck}{
	\Lambda_{MCV}(k) =Im \left(\frac{\omega_{MCV}(k)}{k^2} \right)
	= 
	\begin{cases}
		\frac{1}{2\tau k^2}, &	\text{if} \quad k^2 \geq  \frac{1}{4\tau\lambda},\\
		\frac{1}{2\tau k^2}\left(1\pm \sqrt{1-4\tau\lambda k^2}\right),  & \text{if} \quad k^2 < \frac{1}{4\tau\lambda}.
	\end{cases}
}
On the other hand, if $ k^2 \geq  \frac{1}{4\tau\lambda}$, then the phase velocity has two different values, related to forward and backward propagation:
\eqn{MCV_vph}{
	v_{p}(k) = Re \left(\frac{\omega(k)}{k}\right) = \pm  \frac{\sqrt{4\tau\lambda k^2-1}}{2\tau k},
}
otherwise $v_p=0$. Therefore, in case of $k^2 < 1/(4 \lambda \tau) $  there are no waves, the real part of \re{MCV_drelo} is zero, the amplitude of periodic initial conditions damp exponentially fast. The propagation of the field is purely dissipative.

It is remarkabl{e t}hat in this case, the spectral diffusivity, \re{MCV_drel}, has two different positive values. These values do are not related to forward and backward propagation, they represent two channels of diffusive propagation. When the wave number goes to zero, then the two spectral diffusivities tend to the same expected value, $\lim_{k\to 0}\ehc(k) = \lambda$. 

For the omega dependent representation we get from the dispersion relation \re{MCV_drel}, that
\eqn{MCV_drelk}{
	k^2(\omega)	= \frac{\omega}{\lambda}(\tau \omega - i).
}
Therefore the spectral diffusivity
\eqn{MCV_Lo}{
	\Lambda(\omega) = Im \left(\frac{\omega}{k^2(\omega)}\right) = \lambda.
}
The spectral diffusivity has a single value when expressed as the function of frequency. This example demonstrates well that the wavenumber and angular {velocity-dependent} spectral diffusivities, the temporal and spatial representations express different aspects of dissipation. Seemingly the spatial spectral diffusivity, the definition based on real angular frequency, is more physical for the MCV equation. 


\section{A pure dissipative example: coupled Fourier conductors}

Let us consider a continuum with two temperatures, where rigid Fourier conductors can exchange heat in every point with a constant heat transfer coefficient $\alpha$. This is a pure dissipative system. The system of equations are
\eqn{iebal1_eq}{
	\rho_1 c_1 \partial_t T_1 + \partial_x q_1 &= -\alpha (T_1- T_2), \nl{Fourier1_eq}
	q_1 + \hat\lambda_{1} \partial_x T_1 &= 0, \nl{iebal2_eq}
	\rho_2 c_2 \partial_t T_2 + \partial_x q_2 &= -\alpha (T_2- T_1), \nl{Fourier2_eq}
	q_2 + \hat\lambda_{2} \partial_x T_2 &= 0,
} 
where the two temperatures, heat fluxes and material parameters are denoted by the particular indices.  The wave solution of the equations is 
\eqn{2Fourier_leq}{
	\begin{pmatrix}
		\rho_1 c_1 i \omega + \alpha & - i k  & -\alpha & 0\\
		- i k \hat\lambda_1 & 1 & 0 & 0 \\
		-\alpha & 0 & \rho_2 c_2 i \omega +\alpha & - i k \\
		0 & 0 & - i k \hat\lambda_2 & 1
	\end{pmatrix}
	\begin{pmatrix}T_{10} \\ q_{10} \\ T_{20} \\ q_{20}\end{pmatrix}
= \begin{pmatrix}0 \\ 0 \\ 0 \\ 0\end{pmatrix}.
}
The determinant of the matrix leads to the following dispersion relation
\eqn{2Fourier_drel}{
	\hat\lambda_1\hat\lambda_2 k^4 + \alpha(\hat\lambda_1 + \hat\lambda_2) k^2 - c_1 \rho_1 c_2 \rho_2 \omega^2 + \nnl{l1}
	i \omega\left[(c_1 \rho_1 \hat\lambda_2 + c_2\rho_2\hat\lambda_1) k^2 + \alpha(c_1\rho_1 + c_2\rho_2)\right] = 0.
}
Then the angular frequency is imaginary:
\eqn{2Fourier_dok}{
	\omega_{F2}(k) = \frac{i}{2}\left( \alpha_1 +  \alpha_2 + 
	k^2\left( \lambda_1 + \lambda_2 \right) 
	\pm\sqrt{\left( \alpha_1 -  \alpha_2 + k^2\left( \lambda_1 - \lambda_2 \right)\right)^2 +4\alpha_1\alpha_2}\right).
}

Therefore the temporal spectral diffusivity becomes
\eqn{2Fourier_sdk}{
	\ehc_{F2}(k) &= Im \left(\frac{\omega_{2F}(k)}{k^2}\right)  = -\frac{i \omega_{2F}(k)}{k^2} = \nnl{1p}
	&= \frac{1}{2}\left( \frac{\alpha_1 +  \alpha_2}{k^2} + 
	\lambda_1 + \lambda_2
	\pm\sqrt{\left( \frac{\alpha_1 - \alpha_2}{k^2} + \lambda_1 - \lambda_2 )\right)^2 + \frac{4\alpha_1\alpha_2}{k^4}}\right) .
}

Here $\lambda_1 = \frac{\hat\lambda_1}{\rho_1c_1}$ and $\lambda_2 = \frac{\hat\lambda_2}{\rho_1c_2}$ are the heat diffusivities and 
$\alpha_1 = \frac{\alpha}{\rho_1c_1}$,  $\alpha_2 = \frac{\alpha}{\rho_2c_2}$, respectively.  

The two roots are always real, as it is expected for a pure dissipative continuum. The two heat conduction channels become independent for $\alpha=0$ or in case of infinite wave number. Then the two solutions are $\ehc_1 =\lambda_1$ and $\ehc_2 = \lambda_2$. Therefore it is reasonable to assume that $\ehc_1$ and $\ehc_2$ represent a separation of the coupled heat conduction with two independent heat conduction channels in general, very like a spectral representation of wave propagation. $\ehc_1$ and $\ehc_2$ are the spectral diffusivity components, the diffusivity modes.

The spatial representation is more complicated. The square of the wave number follows as
\eqn{2Fourier_dko}{
	k^2_{F2}(\omega) &=- \frac{1}{2\lambda_1\lambda_2}\left( \alpha_1\lambda_2 +  \alpha_2\lambda_1 + i\omega(\lambda_1+\lambda_2) \right.\nnl{1i}
	&\qquad \left.\pm\sqrt{\left( \alpha_1\lambda_2 -  \alpha_2\lambda_1 + i\omega(\lambda_1-\lambda_2)\right)^2+
		4\lambda_1\lambda_2\alpha_1\alpha_2}\right) =\nnl{2i}
	& = -a + i\omega b \pm \sqrt{c+i\omega d},
}
where 
\eqn{ab}{
	a &= \frac{\lambda_1\alpha_2 + \lambda_2\alpha_1}{2\lambda_1\lambda_2},  \nnl{ab1}
	b &= \frac{\lambda_1 + \lambda_2}{2\lambda_1\lambda_2},					 \nnl{ab2}
	c &= \frac{(\lambda_1\alpha_2 + \lambda_2\alpha_1)^2 - \omega^2 (\lambda_1-\lambda_2)^2}{2\lambda_1\lambda_2},  \nnl{ab3}
	d &= {2\omega(\lambda_1-\lambda_2)  (\lambda_1\alpha_2 - \lambda_2\alpha_1)}{2\lambda_1\lambda_2}.
}   
Then straightforward calculations lead to
\eqn{2Fourier_sdo}{
	\ehc_{F2}(\omega) = Im\left(\frac{\omega}{k(\omega)^2}\right) =\omega \frac{\omega b
		\pm\sqrt{\frac{\sqrt{c^2+(\omega d)^2}-c}{2}}}{	
		\left(-a\pm\sqrt{\frac{\sqrt{c^2+(\omega d)^2}+c}{2}}\right)^2+	
		\left(\omega b\pm\sqrt{\frac{\sqrt{c^2+(\omega d)^2}-c}{2}}\right)^2}
	,
}
The special cases when $\alpha_1=\alpha_2=0$ lead to the expected two heat conduction channels, as in the temporal analysis, $\Lambda_1= \lambda_1$ and $\Lambda_2= \lambda_2$. Also if $\lambda_1= \lambda_2 = \lambda$ one obtains a single solution $\Lambda=\lambda$.


\section{Analysis of damped-dispersive wave propagation: the window condition}

Our last example is related to a system where the wave propagation is unavoidably damped and the conditions of minimal damping are important. This is the case of the window condition of ballistic-diffusive wave propagation of the 9-field theory of Extended Thermodynamics, which is obtained from the moment series expansion of kinetic theory \cite{MulRug98b}. In this case we analyse the following system of equations: 
\eqn{iebalm_eq}{
	\partial_t e + \cc^2 \partial_x p &= 0, \nl{qconst_eq}
	\partial_t p+ \frac{1}{3}\partial_x e  + \partial_x N + \frac{p}{\tau_R}&= 0,\nl{Qconst_eq}
	\partial_t N + \frac{4}{15} \cc^2 \partial_x p + \left(\frac{1}{\tau_R}+ \frac{1}{\tau_N}\right) N&= = 0.
} 
Here the tree equations re the balances of the density of the internal energy, $e$, the heat flux, $\cc^2 p$, and the flux of the heat flux $N$. $\tau_R$ and $\tau_N$ are the relaxation times of the R(resistive) and N (normal) processes, $\cc$ is the Debye speed of the phonons in the crystal. Then the transfer matrix is 
\eqn{difbal_leq}{
	\begin{pmatrix}
		i \omega & - \cc^2 i k  & 0\\
		- \frac{i k}{3} & i \omega + \frac{1}{\tau_R} & - i k \\
		0 & -\frac{4ik}{15} \cc^2 & i \omega + \frac{1}{\tau_R}+\frac{1}{\tau_N} 
\end{pmatrix}}
The determinant of the matrix leads to the following dispersion relations
\eqn{difbal_drel}{
	- i \omega^3 -\omega^2 \left(\frac{1}{\tau}+\frac{1}{\tau_R}\right) +i \omega \left(\frac{1}{\tau_R\tau} + \frac{3}{5} \cc^2 k^2\right) + \frac{\cc^2}{3 \tau}k^2 =0
}
where $\tau = \frac{\tau_R\tau_N}{\tau_R+\tau_N}$. The temporal dispersion relation is a solution of a third order polinomial, but straightforward calculation leads to the spatial form of the spectral diffusivity $\ehc(\omega)$ function
\eqn{ehc_9F}{
	\ehc_{9F}(\omega) = \frac{\cc^2\tau_R}{15}\frac{5 +  \omega^2\tau\left(4\tau_R+9\tau\right)}{(1+\omega^2\tau_R^2)(1+\omega^2\tau^2)}
}
It is a monotonous decreasing function of the angular frequency $\omega$, where the two limiting cases are 
\eqn{limehc_9F}{
	\lim_{\omega\to 0}\ehc_{9F}(\omega) = \frac{\cc^2}{3}\tau_R, \qquad
	\lim_{\omega\to \infty}\ehc_{9F}(\omega) = 0. 
} 
The maximum of $\ehc_{9F}(\omega)$ characterises minimum damping and dispersion in the language of wave propagation concepts. It is easy to calculate, that the above function is monotonically decreasing (for positive arguments) if $\frac{\tau}{\tau_R} > \frac{\sqrt{6}-1}{2} \approx 0.72$ and has a single maximum otherwise. This maximum will be at the following frequency
\eqn{wc}{
\omega_{wincon}^2 = \Omega_0^2+\sqrt{ \Omega_0^2\left( \Omega_0^2-\frac{\Omega_2^4}{\Omega_1^2}\right)+ \Omega_2^4},
}
where
\eqn{omdef}{
 \Omega_0^2 = \frac{5}{\tau(9\tau+4\tau_R)}, \quad
 \Omega_1^2 = \frac{1}{\tau^2+\tau_R^2}, \quad
 \Omega_2^2 = \frac{1}{\tau\tau_R}.
}

The window condition  is obtained if we approximate this solution,  originally with the following assumptions $1 \gg \tau_N\omega$ and $1 \ll \tau_R\omega$, according to \cite{GuyKru64a,MulRug98b}. Then 
\eqn{ehc_9Fa}{
	\ehc_{9F}(\omega) \approx \frac{\cc^2}{3}\tau_R\left(1 + \frac{4}{5}\tau_N\tau_R\omega^2\right).
}

However, the existence of maximal spectral diffusivity does not depend on this approximation.

\section{Summary and discussion}

The concept of spectral diffusivity is suggested for analysing dissipation properties of continua. It is calculated from a dispersion relation and defined by $\ehc(\omega) := Im \left( \frac{ \omega}{k(\omega)^2}\right)$ and $\ehc(k) :=Im \left(\frac{ \omega(k)}{k^2}\right)$, for a temporal and spatial characterisation of diffusivity. 
It is a tool for a modal analysis for damping, as the phase speed for the ideal, nondissipative wave propagation. We have shown that in the case of Fourier heat conduction, it is equal to the thermal diffusivity. We have also investigated a simple combination with wave propagation and a pure dissipative coupled heat conduction. It was demonstrated that the concept could give an insight into the structure of dissipation, identifying separate diffusion channels.  We have also investigated the {9-field} equations, the best-known theory of second sound, and we have recovered that maximal spectral diffusivity corresponds to the celebrated window condition as a first approximation.

$\ehc(\omega)$ and $\ehc(k)$ are not the same. The difference of the phase velocities $v_p(k)$ and $v_p(\omega)$, defined in \re{wavesol1} and in \re{wavesol2} is similar and originated in the mathematical fact that  the complex functions $\omega(k)$ and $k(\omega)$ are inverses. The wave number dependent definition of phase velocity $v_p(k) = Re(\omega(k)/k)$ is more accepted than the alternative definition $v_p(\omega) = Re(\omega/k(\omega))$. The later one is a discussed topic in seismology, it is well interpreted only in the long wavelength $k=0$ limit \cite{Fut62a}.


In our case, for spectral diffusivity, the wavenumber dependence can reflect the influence of material heterogeneities. This hypothesis was demonstrated for effective heat conduction coefficient in superlattices \cite{VazEta20a}.

\section{Acknowledgement}   
The research reported in this paper and carried out at BME has been supported by the grants National Research,
Development and Innovation Office-NKFIH FK 134277, K 124366, and by the NRDI Fund (TKP2020 NC, Grant
No. BME-NCS) based on the charter of bolster issued by the NRDI Office under the auspices of the Ministry for
Innovation and Technology. This paper was supported by the J\'anos Bolyai Research Scholarship of the Hungarian
Academy of Sciences (R. K.). 

\bibliographystyle{unsrt}

\end{document}